\documentclass[%
 reprint,
 amsmath,amssymb,
 aps,
]{revtex4-1}

\usepackage{graphicx}
\usepackage{dcolumn}
\usepackage{bm}

\usepackage{balance}
\usepackage{times,mathptmx}
\usepackage{lastpage}
\usepackage{float}
\usepackage{fancyhdr}
\usepackage{fnpos}
\usepackage[english]{babel}
\usepackage{array}
\usepackage{droidsans}
\usepackage{charter}
\usepackage[utf8]{inputenc}
\usepackage[usenames,dvipsnames]{xcolor}
\usepackage{setspace}
\usepackage[compact]{titlesec}

\usepackage{fixltx2e}
\usepackage{siunitx}
\usepackage{amssymb}

\usepackage{subfigure}

\usepackage{systeme}

\usepackage{cleveref}
\crefname{equation}{eq.}{eqs.}
\Crefname{equation}{Eq.}{Eqs.}

\crefname{figure}{fig.}{figs.}
\Crefname{figure}{Fig.}{Figs.}

\usepackage{epstopdf}

\definecolor{cream}{RGB}{222,217,201}
\graphicspath{ {./pics/} }
\DeclareGraphicsExtensions{.jpg,.jpeg,.png,.pdf,.tiff}

\begin{document}


\title{Progressive friction mobilization and\\enhanced Janssen's screening in confined granular rafts}

\author{Oscar Saavedra V.}
\affiliation{Departamento de F\'{\i}sica, Universidad de Santiago de Chile, Av. Ecuador 3493, Santiago, Chile}
\author{Herv\'e Elettro}%
 \email{herve.elettro@usach.cl}
 \homepage{herveelettro.wordpress.com}
\affiliation{Departamento de F\'{\i}sica, Universidad de Santiago de Chile, Av. Ecuador 3493, Santiago, Chile}%


\author{Francisco Melo}
\email{francisco.melo@usach.cl}
\affiliation{Departamento de F\'{\i}sica, Universidad de Santiago de Chile, Av. Ecuador 3493, Santiago, Chile}%


\date{\today}

\begin{abstract}
Confined two dimensional assemblies of floating particles, known as granular rafts, are prone to develop a highly non linear response under compression. 
Here we investigate the transition to the friction-dominated jammed state and map the gradual development of the internal stress profile with flexible pressure sensors distributed along the raft surface. Surprisingly, we observe that the surface stress screening builds up much more slowly than previously thought, and that the typical screening distance later dramatically decreases. We explain this behaviour in terms of progressive friction mobilization, where the full amplitude of the frictional forces is only reached after a macroscopic local displacement.
At further stages of compression, rafts of large length-to-width aspect ratio experience much stronger screenings than the full mobilization limit described by the Janssen's model. We solve this paradox using a simple mathematical analysis and show that such enhanced screening can be attributed to a localized compaction front, essentially shielding the far field from compressive stresses.

\end{abstract}

\maketitle


Granular rafts are self-assembled structures of floating particles at a fluid-fluid interface. They offer a simple realization of a two dimensional athermal system that combines capillary and granular properties \citep{Abkarian2013,Alarcon2012,Cicuta2009,Cicuta2003,Saingier2017,Huang2005,Brunet2008,Mitarai2006}. 
Mechanics of particles at liquid interfaces is of practical importance to a broad range of systems, from self cleaning surfaces to industrial processes \citep{Strauch2012,Lafuma2003,Chevalier2013,Dai1999}.
In particular, particle-laden interfaces are relevant to particle-stabilized foams \citep{Tambe1994,Gonzenbach2006,Binks2006,Du2003} and 
for several innovative applications such as non-wetting liquid marbles 
\citep{Taccoen2016,Aussillous2001, mchale2015, Wong2017}, relying on both the solid-like properties of particles 
and the stabilizing effect of fluid surface tension \citep{lagubeau_melo,Protiere2017}.

%
In confined granular materials, the pressure field saturates under compaction due to frictional interactions with the container's walls, as first observed by Janssen \citep{Janssen1895} in silos. Screening occurs on a typical length scale $\lambda_\text{j} = \tfrac{1}{2\mu_\text{j}\nu}$ in units of channel width W, where $\mu_\text{j}$ is the friction coefficient of the particle-wall contact and $\nu$ is the Poisson's ratio of the granular raft. 
\citet{Cicuta2009} showed that stress transmission appears to be screened over a similar length scale 
in granular rafts. 
However, Janssen's state manifests only 
when frictional forces are fully mobilized, at the limit of Coulomb's cone for sliding onset. 

Here we show that the internal stress distribution of confined granular rafts does not corroborate the classical Janssen's hypothesis,  
and instead reveals that regions of different screening behaviours may coexist.

Using local pressure sensors, we explore the rafts internal mechanics and show that the instantaneous screening length swiftly decreases as frictional forces are progressively mobilized. However, the final characteristic screening length scale is found to be dramatically smaller than Janssen's prediction.
We relate this enhanced screening to the presence of a compaction front, where a gradient of packing fraction develops and shields the far field from compressive stresses. 
%

We develop a simple theoretical framework to offer a revision of the classical Janssen's model, and account for both the progressive mobilization of frictional forces and the effect of a gradient of packing fraction. 
Our analysis indicates that the screening length scale is significantly affected by the local elastic response of the raft, which bias the interpretation of experimental data such as the force transmitted through the raft.

%

Our experimental cell consists of four barriers made of Teflon, three of which are fixed and one left mobile, forming a rectangular channel of controllable aspect ratio. All experiments have been repeated at fixed channel width of $10$ and $25$\si{\milli\meter}, much larger than the submillimetric particles. 
Gaps are allowed on each side of the moving barrier to prevent barrier-wall friction. Additionnally, the gaps are finely adjusted with a 3-axis translation stage to be smaller (about $100$\si{\micro\meter}) than the particle size, in order to avoid leaking. 
A motorized microstage provides controlled displacement 
and sensors of $10\mu$N sensitivity register the force acting on the moving and end barriers (see Supplemental Information \citep{suppmat}). 


The trough is filled with deionized water up to the rim of the Teflon walls, to obtain flat meniscii and prevent any influence of gravitational pulls, inward or outward. Compression is applied at 
$100$ \si[per-mode=symbol]{\micro\meter\per\second} until the onset of the buckling instability. No dependence on the compression rate was observed. We use glass particles that are made hydrophobic by means of surface treatment (silanization with \textsuperscript{1}H,\textsuperscript{1}H,\textsuperscript{2}H,\textsuperscript{2}H-perfluorooctyltriethoxysilane (C\textsubscript{14}H\textsubscript{19}F\textsubscript{13}O\textsubscript{3}Si), see Supplemental Material \citep{suppmat}). We expect an equilibrium contact angle of $110^\circ$ \citep{Wang2011}. To prevent crystallization, particles are polydisperse, 
ranging from $0.2$\si{\milli\meter} to $0.3$\si{\milli\meter} in size. The maximum achievable packing fraction is found to be $\phi_\text{max} = 0.848 \pm 0.04$. The initial packing fraction is set to $\phi_0 = 0.652 \pm 0.03$ for a controlled reference state for all presented experiments. $\phi_0$ corresponds to a loose state for the particle raft and allows the recording of the entire process of mobilization of the frictional forces.

Since the initial packing fraction $\phi_0$ 
is relatively low compared to the random close packing fraction for dry particles $\phi_\text{rcp} \simeq 0.775$ \citep{Meyer2010}, the raft undergoes important rearrangement during compression \citep{Campbell2006}. 
%

\begin{figure}[ht]
\begin{center}
\includegraphics[width=\columnwidth]{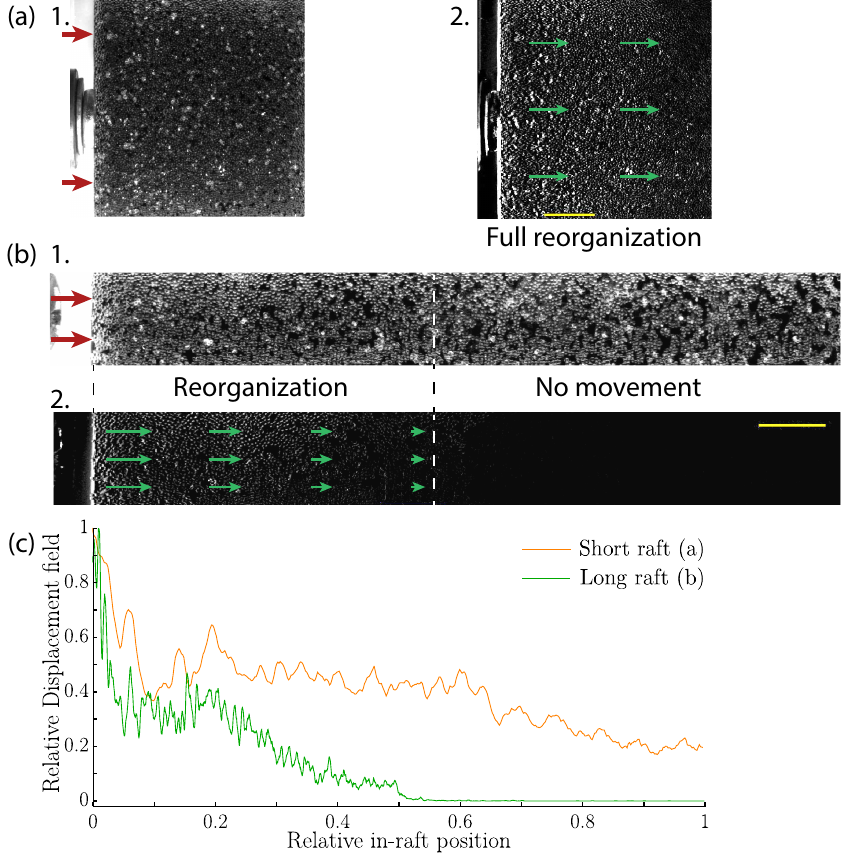}
\caption{Optical images (1) are correlated to reveal the local deformation field as bright and dark spots 
(2) (see Supplemental Materials \citep{suppmat}). For small rafts the deformation field is homogeneous (a- top), whereas for large rafts the deformation is quickly localized (b- bottom and Supplemental Videos \citep{suppmat}) and an untouched tail zone develops. The compaction front spreads over a lengthscale $\lambda_\phi \simeq 4.2 \pm 0.5$ (measured optically in width $W$ units). 
Red arrows represent imposed displacement and 
thin green arrows show the local grain displacements (not to scale). 
(c) Normalized local grain displacement profile versus in-raft position for short and long rafts. The scale bars represent $5$ \si{\milli\meter}.}
\label{compaction}
\end{center}
\end{figure}

For rafts shorter than the Janssen's screening length $\lambda_\text{j}$, the stress distribution in the raft is nearly homogeneous. Grain displacement thus decreases linearly from the imposed displacement $\Delta L=L_0-L$ at the moving barrier towards zero at the fixed barrier, indicating a homogeneous compaction (\cref{compaction}(a)). We note $L_0$ the initial length of the granular raft and $L$ its instantaneous length. However, image correlation reveals that long rafts have grain displacements localized near the compression barrier (\cref{compaction}(b) and Supplemental Material \citep{suppmat}). 

%

\begin{figure}[ht]
\begin{center}
\includegraphics[width=\columnwidth]{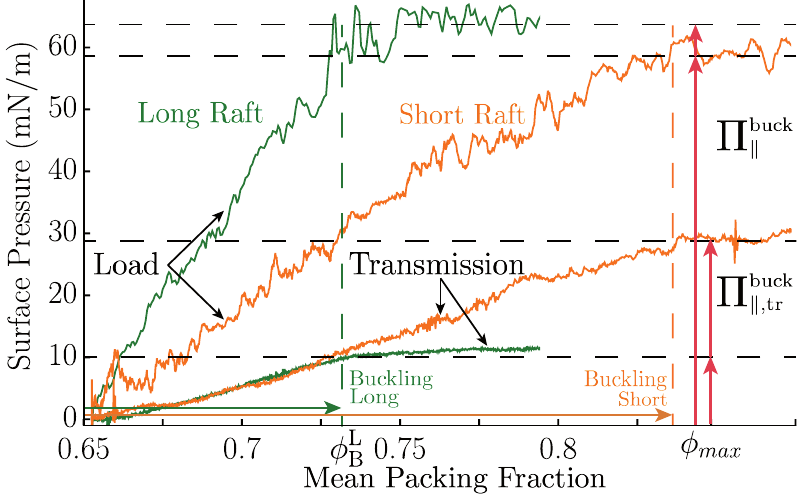}
\caption{Typical loading (top) and transmitted (bottom) surface pressures of rafts of aspect ratio at buckling $2.1 \pm 0.1$ (orange curves) and $5.2 \pm 0.2$ (green curves) versus mean packing fraction $\phi_0 \tfrac{L_0}{L_0-\Delta L}$.  
The uncompacted tail zone of the long raft leads to early buckling (green) at lower mean packing fraction (here $\phi_\text{B}^\text{L} \simeq 0.73 < \phi_\text{max}$). The (homogeneous) short raft follows its initial path (orange) up to a mean packing fraction close to $\phi_\text{max}$. The loading (resp. transmitted) buckling surface pressure $\Pi_{\parallel}^{\text{buck}}$ (resp. $\Pi_{\parallel\text{, tr}}^{\text{buck}}$) are defined on the right hand side.}
\label{caracteristica}
\end{center}
\end{figure}

The transmitted force increases progressively under compression (\cref{caracteristica}). Further compression leads to buckling, where the raft develops an out-of-plane wrinkled structure. 
Long rafts reach buckling at lower mean packing fraction (see \cref{caracteristica}), due to the presence of the finite penetration length for compaction (see \cref{compaction}). 
We measured optically the compaction front size $\lambda_\phi$ as the distance from the moving barrier to the point where the intensity of the grain displacement drops below background noise level. This yields $\lambda_\phi \simeq 4.2 \pm 0.5$ in width W units.
Direct optical measurements of the packing fraction confirm the value of $\lambda_\phi$ 
(see Supplemental Material \citep{suppmat}).
%

%

%
%
We follow \citet{Cicuta2009} and measure the maximum transmitted surface pressure ($\Pi = F/W$) along the raft main axis $\Pi_{\parallel\text{,tr}}^{\text{buck}}$ at the buckling threshold, upper bound for the applied force (see \cref{caracteristica}).

The initiation of the transition to solid state should occur at the random loose packing 
limit $\phi_\text{rlp}$, where a connected network of contacts allows transmission of forces throughout the entire granular medium \citep{Thornton1997}. For particles with a purely repulsive (steric) interaction at contact, geometrical arguments suggest $\phi_\text{rlp} \simeq 0.775$ \citep{Meyer2010}. However, the attractive capillary interactions between floating particles force earlier contact and allow transmission at a lower packing fraction \citep{Berhanu2010}, from the beginning of the compression in our case.
%
%
Note that in the initial state, the raft self-organizes into clusters (see \cref{compaction}). This is caused by an effective particle-particle attraction potential \citep{Waitukaitis2011}, due to capillary meniscii between the particles \citep{Vella2005,Stamou2000}. The resulting void structure has a well defined wavelength, similar to the cracks in a stretched cohesive granular layer, which are induced by the softening of the capillary meniscii with increasing interparticle distance \citep{Alarcon2010}.
Although locally inhomogeneous, the resulting granular rafts have an homogeneous concentration of clusters all throughout. 
Stirring before each experiment ensures that the system reaches a stress-free configuration at the desired packing fraction. Hence the buckling threshold does not depend on neither the initial state nor the chosen initial packing fraction for homogeneous rafts.

\begin{figure}[ht]
\begin{center}
\includegraphics[width=\columnwidth]{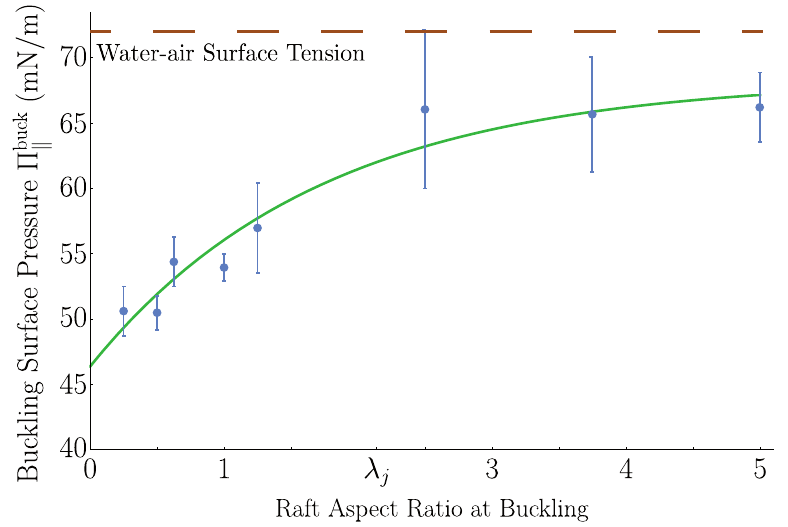}
\caption{The buckling surface pressure of particle rafts increases with its aspect ratio at buckling. This is in contrast to the assumption of \citet{Cicuta2009} that the buckling surface pressure $\Pi_{\parallel}^{\text{buck}} = \gamma_{wo}$, which is recovered only for rafts of very large aspect ratio.}
\label{buckling}
\end{center}
\end{figure}

In order to properly evaluate the transmission factor $\Pi_{\parallel\text{,tr}}^{\text{buck}}/\Pi_{\parallel}^{\text{buck}}$, we measure the reference loading surface pressure at buckling $\Pi_{\parallel}^{\text{buck}}$. Interestingly, we find that the buckling threshold depends on the raft aspect ratio with a $30\%$ variation over the range of study, see \cref{buckling}. This is in contrast to the assumption of constant buckling load of \citet{Cicuta2009}. We fit the experimental measurements with a saturating exponential in the form $\Pi_{\parallel}^{\text{buck}} = \Pi_{\parallel, 0}^{\text{buck}} + (\Pi_{\parallel, \infty}^{\text{buck}}-\Pi_{\parallel, 0}^{\text{buck}})(1-\exp(-AR/\lambda_\text{load}^\text{buck} ))$ with $\lambda_\text{load}^\text{buck}$ the saturation length scale, $\Pi_{\parallel, 0}^{\text{buck}}$ (resp. $\Pi_{\parallel, \infty}^{\text{buck}}$) the surface pressure at buckling for an infinitely short (resp. long) raft and $AR$ the aspect ratio at buckling of the raft, defined as the ratio of the instantaneous length to width of the raft. The best fit 
yields a saturation length scale $\lambda_\text{load}^\text{buck} = 2.59 \pm 0.8$. This is compatible with Janssen's prediction $\lambda_\text{j} = \tfrac{1}{2\mu_\text{j}\nu} = 2.17 \pm 0.13$, where $\mu_\text{j} = 0.401 \pm 0.024$ (measured by plane inclination) and $\nu = \tfrac{1}{\sqrt{3}}$ 
\citep{Vella2004,Cicuta2009}. We also find $\Pi_{\parallel, 0}^{\text{buck}} = 47 \pm 1.5$\si[per-mode=symbol]{\milli\newton\per\meter} and $\Pi_{\parallel, \infty}^{\text{buck}} = 67 \pm 4.5$\si[per-mode=symbol]{\milli\newton\per\meter}. Note that only $\Pi_{\parallel, \infty}^{\text{buck}}$ is close to the value of the surface tension of the water-air interface $\gamma_{wo} = 72$\si[per-mode=symbol]{\milli\newton\per\meter}, which is the reference value taken by \citet{Cicuta2009}.
We interpret this increase of buckling onset as a consequence of the interplay between the compaction front and stresses at the walls.

%
%

\begin{figure}[ht]
\begin{center}
\includegraphics[width=\columnwidth]{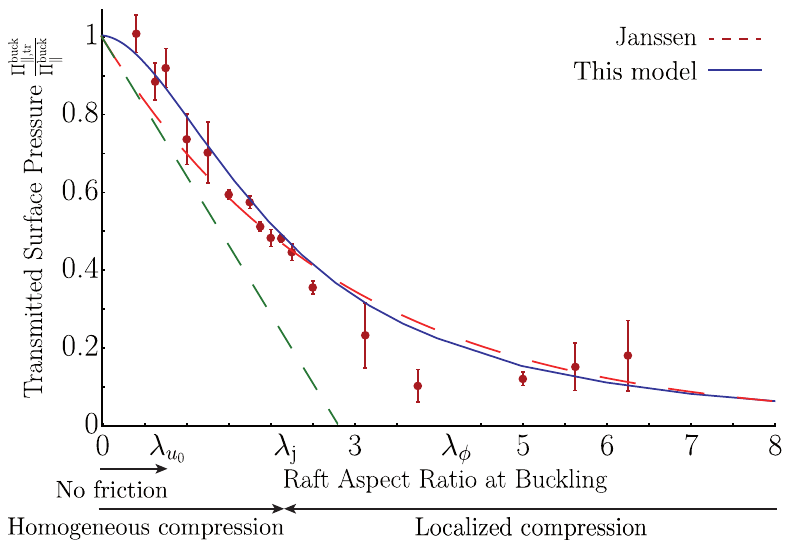}
\caption{Friction-induced decay of the transmitted surface pressure $\Pi_{\parallel\text{,tr}}^{\text{buck}}/\Pi_{\parallel}^{\text{buck}}$ for rafts of increasing aspect ratio. 
The dashed (red) line corresponds to the best experimental fit with constant screening length. 
The dashed (green) line 
points to the measured value of the screening length $\lambda_\text{j}^\text{tr}$.
The solid (blue) line shows the numerical solution of the present model. Note that although the mechanisms in our model (progressive friction mobilization and later enhanced screening) are conceptually different from Janssen's model (fully mobilized homogeneous friction), the transmitted surface pressure profiles are strikingly similar. This calls for accurate measurements of the internal stress profile (see \cref{profile}), necessary to discriminate the mechanisms at play.}
\label{transmission}
\end{center}
\end{figure}

We then measure experimentally the transmitted surface pressure, which is presented in \cref{transmission}.
In agreement with \citet{Cicuta2009} we observe that an exponential 
screening with length scale $\lambda_\text{j}^\text{tr} \simeq \lambda_\text{j}$ fits reasonably well experimental data (\cref{transmission}, dashed red line), even though the measurement leads to invariably overestimated screening lengths (further discussed below, see green dashed line on \cref{transmission}). 
In addition, force transmission is almost total for short rafts, and decreases noticeably faster than Janssen's prediction for rafts of intermediate aspect ratio. This suggests that friction mobilization is only partial 
for short rafts and that screening is enhanced for intermediate rafts.

\begin{figure}[h!t]
\begin{center}
\includegraphics[width=\columnwidth]{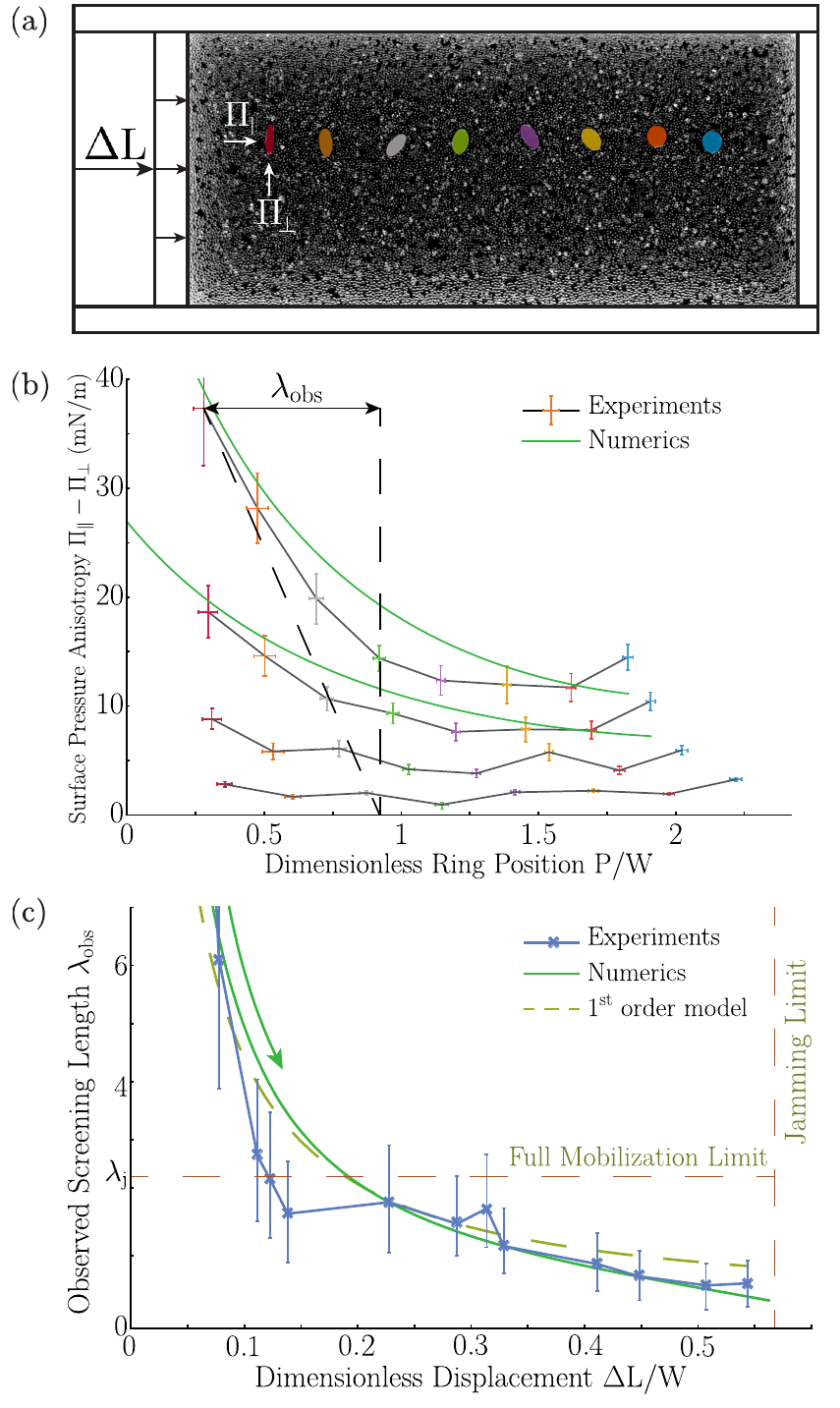}
\caption{(a) View of a raft of initial aspect ratio $2.6$ under $54\%$ surface pressure (normalized by the buckling threshold) with 8 rings sensors. Elliptical masks are adjusted for quantitative measurements. 
(b) Rings deformations reveal progressive development of exponential profiles. Profiles correspond to the following normalized surface pressures : $9\%, 36\%, 64\%, 82\%$ (from bottom to top). Numerical predictions (solid green) 
at corresponding displacement $\Delta L$ show good agreement with experimental data near the compression wall. We observe descrepancies near the fixed end wall, possibly related to residual 2D effects. The colours of the experimental points refer to the corresponding ring in (a).
(c) Observed screening length $\lambda_\text{obs}$ based on the pressure field profile measured near the compression wall. 
Experimental data (blue crosses), numerical simulations (green) and first order approximation (dashed yellow) show good agreement. Note that the observed screening length $\lambda_\text{obs}$ can reach values considerably smaller than the Janssen's case 
(horizontal brown dashed line).}
\label{profile}
\end{center}
\end{figure}

Hence, we develop flexible ring sensors to map the onset of friction mobilization 
directly from within the granular raft (see Supplemental Material \citep{suppmat}). 
We use a raft of intermediate initial aspect ratio $2.6$ (and $1.8 \lesssim \lambda_\text{j} $ at buckling) with 8 rings spread homogeneously. 
Image correlation indicates that the presence of the rings does not significantly change the deformation field at large scale. 
Further analysis shows 
a nearly homogeneous packing fraction.
The distance between two rings is set to at least $5$\si{\milli\meter} (roughly $20$ particles) to prevent coupling between rings as inclusions (\cref{profile}(a)) \citep{Eshelby1957}.
Calibration loading tests 
show that rings respond to $\Pi_{\parallel}-\Pi_{\perp}$ (defined with reference to the compression axis) by taking an elliptical shape, which we measure optically 
(see Supplemental Materials \citep{suppmat}).
Significant fluctuations of both the orientation and eccentricity of rings are observed, 
due to the heterogeneous nature of the force network. We average 33 independent realizations to improve statistics. At small load, $\Pi_{\parallel}-\Pi_{\perp}$ is found to be nearly independent upon the distance from the compression barrier ($z$). For higher loads, a zone of stress localization gradually develops, as observed on \cref{profile}.
%
%
To characterize the progression of this zone, we fit an exponential law with variable screening length $\lambda_\text{obs} (\Delta L)$ to the profile near the compression barrier. $\lambda_\text{obs}$  decreases rapidly with  $\Delta L$, and reaches a value significantly smaller than the Janssen's prediction $\lambda_\text{j}$ (see \cref{profile}(c)). This suggests the presence of an additional mechanism contributing to screening, which is discussed in the following model. 
%

\label{Poisson_exp}
We also determine the raft's effective Poisson's ratio $\nu$, knowing that the 
ring sensors measure the surface pressure anisotropy $\Pi_\parallel - \Pi_\perp$, while the end wall sensor measures 
$\Pi_\parallel$. Recalling that $\Pi_\perp = \nu \Pi_\parallel$ \citep{Cicuta2009}, we obtain $(\Pi_\parallel - \Pi_\perp)/\Pi_\parallel = 1 - \nu$.
%
Our measurements show that $\nu$ does not vary significantly within the considered packing fraction range (see Supplemental Materials \citep{suppmat}), taking values from $0.52$ to $0.7$ with an $0.09$ error. 
Thus the fact that $\lambda_\text{obs}$ is significantly smaller than the expected screening length $\lambda_\text{j}$ cannot be attributed to a progressive variation of $\nu$. 
%
%
%
%
\label{model}
%
In the following we develop a simple 1D framework for the analysis of our experimental results.

We first write a force balance in a band of width $dz$ perpendicular to the raft main axis 
, $w(\Pi_\parallel(z)-\Pi_\parallel(z+dz)) - 2\mu_\text{obs}\Pi_\perp(z)dz = 0$. 
%
We assume that i) surface stresses are linked through $\Pi_\perp(z)=\nu\Pi_\parallel (z)$, ii) the raft responds linearly \citep{Jiang2003} to compression with an effective elastic modulus $E$, and iii) lateral walls prohibit transverse spreading, such that $\Pi_\parallel = 2R\tfrac{E}{1-\nu^2} \tfrac{d u_z}{d z}$, where 
$u_z$ is the displacement field along the raft main axis and $R$ is the radius of the particles.

It is important to emphasize that Janssen's exponential screening is theoretically an upper bound, reached only for total friction mobilization everywhere. 
Describing friction mobilization raises fundamental questions difficult to address without a detailed knowledge of the rheology of granular rafts. Therefore we make additional simplifying assumptions: i) frictional forces at walls are not mobilized in the initial state, ii) progressive friction mobilization occurs and depends solely on the relative displacement of grains at walls. iii) total friction mobilization is achieved for a typical length scale of displacements $u_0$, which depends on the initial packing fraction $\phi_0$.  Following \citet{Vivanco2016}, we write $\mu_\text{obs} = \mu_\text{j}(1-e^{-\tfrac{u_z}{u_0}})$, which ultimately leads to

\begin{equation}
\label{force_balance_tot}
\frac{d^2 u_z}{d z^2}= -\dfrac{1-e^{-\tfrac{u_z}{u_0}}}{\lambda_\text{j}}\frac{d u_z}{d z}
\end{equation}

We emphasize that the proposed model is an extension of the Janssen's model, with the addition of a new parameter, the mobilization length scale for the frictional forces $u_0$. The classical Janssen's screening is recovered in the limit of vanishing mobilization length scale, $u_0 \to 0$, i.e.\ always fully mobilized friction.  

Although \cref{force_balance_tot} does not depend on the elastic modulus $E$, a reliable expression of this quantity is required to convert grain displacement into surface pressure.  We  model the effective elastic modulus of the granular raft as,

\begin{equation}
\label{Young}
E\left(\phi\right) \propto \frac{\phi - \phi_\text{c}}{\phi_\text{max} - \phi_\text{c}}\times\frac{1-\nu}{1-\phi}\frac{\gamma}{2R} \text{ if } \phi > \phi_\text{c}
\end{equation}

\noindent where $\gamma$ is the liquid surface tension, $\phi_\text{c}$ the random loose packing
threshold and $\phi_\text{max}$ the maximum packing fraction. This form of the elastic modulus is a simple extension of the near buckling prediction of \citet{Vella2004} that recovers vanishing elastic modulus 
below $\phi_\text{c}$. For an initially loose assembly of floating particles, $\phi_\text{c} = 0.44\pm 0.02$ \citep{Berhanu2010}, significantly lower than the random loose packing fraction for dry particles $\phi_\text{rlp} \simeq 0.775$, due to attractive capillary forces.
Although the prediction 
of \citet{Vella2004} qualitatively recovers the following results, we found this new estimate to be more accurate. 

%
Finally, the surface strain yields
\begin{equation}
\label{packing}
\phi\left(z,\Delta L\right) = \phi_0\left(1-\dfrac{d u_z}{d z}\right)
\end{equation}

We then compute the system of equations formed by \cref{force_balance_tot,Young,packing} 
for increasing $\Delta L$ under the boundary conditions 
$u_z(0) = \Delta L$ and $u_z(L_0-\Delta L) = 0$, with $z=0$
We stop calculations when the packing fraction $\phi$ at the compression barrier reaches $\phi_\text{max}$, a natural condition for the buckling onset. 
The results of the numerical simulations are confronted to the experimental data and the Janssen's predictions, and discussed in the following section.
%
%
%
%

\label{discussion}
As a check of consistency of our approach, 
we use \cref{packing} to predict the value of the buckling load. We obtain $\Pi_\parallel^{\text{buck}} \simeq \gamma\times\tfrac{\phi_\text{max}-\phi_0}{\phi_0(1-\phi_\text{max})(1+\nu)}\simeq 50-100$\si[per-mode=symbol]{\milli\newton\per\meter}, consistent with experimental values. 
Assuming friction is always mobilized ($u_0 \to 0$, Janssen's case), the best fit of the experimental data yields a measured screening length $\lambda_\text{j}^\text{tr} = 2.82 \pm 0.09$ (\cref{transmission}), higher than the Janssen's case $\lambda_\text{j} = \tfrac{1}{2\mu_\text{j}\nu} = 2.17 \pm 0.13$.
%
%
This discrepancy is explained by partial friction mobilization, especially important for small rafts. This hypothesis is further sustained by the compelling difference between the screening length directly obtained from internal stress measurements $\lambda_\text{obs}$ with respect to $\lambda_\text{j}$ (see \cref{profile}(c)). This can be related qualitatively with the study of \Citet{Boutreux1997}, who also found theoretical evidences of screening lengths $30\%$ higher than Janssen's prediction after a pressure step propagated inside a granular column.

To better grasp the complex behaviour of the coupled system of \cref{force_balance_tot,Young,packing}, we write a first order approximation by developing the displacement field $u_z$ near the compression barrier as $u_z(z) \approx \Delta L - \epsilon(z)$ (with $\epsilon \ll \Delta L$). 
\Cref{force_balance_tot} thus yields 
\begin{equation}
\label{epsilon}
\frac{d^2 \epsilon}{d z^2}= -\dfrac{1-e^{-\tfrac{\Delta L}{u_0}}}{\lambda_\text{j}}\frac{d \epsilon}{d z}
\end{equation}

and the related observed screening length for the displacement field

\begin{equation}
\label{lambda_obs_u}
\lambda_\text{obs}^u = \dfrac{\lambda_\text{j}}{1-e^{-\tfrac{\Delta L}{u_0}}}.
\end{equation}

(see \cref{packing}).
Injecting $u_z(z) = \Delta L\phantom{.}\exp(-z/\lambda_\text{obs}^u)$ into \cref{packing} shows that the packing fraction field $\phi(z)$ is screened over the same length scale than the grain displacement.
However, we expect a different screening length for the surface pressure field $\Pi_\parallel$, since it is the combination of the elastic modulus field $E(z)$ and the displacement field $u_z (z)$ in $\Pi_\parallel(z) \propto E(z) \tfrac{d u_z}{d z}(z)$. We emphasize that this separation of screening lengths is a consequence of the observed gradient of compaction, and is unique to the present model. This represents a fundamental difference with the Janssen's model, where all physical parameters are screened over the same lengthscale. Using \cref{Young}, we obtain the observed screening length for the surface pressure field $\lambda_\text{obs}^\Pi$ and its relation to 
$\lambda_\text{obs}^u$ as,
%

\begin{equation}
\label{lambda_obs_Pi}
\lambda_\text{obs}^\Pi = \frac{\lambda_\text{obs}^u}{1 + \left(\tfrac{\phi_0}{1-\phi_0} + \tfrac{\phi_0}{\phi_0-\phi_\text{c}}\right) \frac{\Delta L}{\lambda_\text{obs}^u}}
\end{equation}

\noindent where the coefficient $a(\phi_0,\phi_\text{c}) =\tfrac{\phi_0}{1-\phi_0} + \tfrac{\phi_0}{\phi_0-\phi_\text{c}} \simeq 5.1 \pm 0.3$.

$\lambda_\text{obs}^\Pi$ decreases 
sharply until $\Delta L \simeq u_0$ and then further decreases like $1/\Delta L$ for $\Delta L \gtrsim \lambda_\text{j}/a(\phi_0,\phi_\text{c})$ (see \cref{lambda_obs_Pi,lambda_obs_u}). At buckling, $\lambda_\text{obs}^\Pi$ reaches $\lambda_\infty \simeq \lambda_\text{j}/3 \ll \lambda_\text{j}$ for the raft of initial aspect ratio $2.6$ presented in \cref{profile}. 
This shows that screenings stronger that the Janssen's fully mobilized limit are possible. 
This enhanced screening process 
may explain the giant overshoot effect observed by
\citet{Ovarlez2003} in loaded vertical granular columns, where a zone of finite extension is preferentially restructured, leading to the same type of gradient of elastic modulus. 
\Cref{profile}(b) also shows a breakdown of the exponential-like surface pressure profile near the fixed end wall (plateauing and even reincrease). We interpret this phenomenon in terms of 2D effects: the non-penetration condition at the end wall enhances lateral spreading of the particles and produces backwards recirculation.

Our numerical study 
and its first order approximation $\Pi_\parallel(z) \propto \exp{\left(-z/\lambda_\text{obs}^\Pi (\Delta L)\right)}$ show good agreement with experimental data for $u_0 = 0.22 \pm 0.1$ in width W units, see \cref{profile}(b,c).  
The mobilization length $u_0$ is thus on the order of a few millimeters, i.e. very large with respect to the particle diameter and at least 3 orders of magnitude above values for dry particles in a silo configuration
\citep{Vivanco2016}. We relate this exceptionally high value to the possibility of local 2D reorganization near the walls at low packing fraction. A natural extension of this study would be to measure the mobilization length $u_0$ at varying initial packing fraction $\phi_0$. We expect to recover the Janssen's model close to the jamming transition, that is $u_0 \to 0$ for $\phi_0 \to \phi_\text{max}$.
%



%
%
%

%
%
We observe three regimes depending on the raft aspect ratio at buckling $L_\text{Buck}/W$.
For small rafts, friction is never fully mobilized
, since buckling occurs for displacements smaller than $u_0$. We call $\lambda_{u_0}$ the limit aspect ratio at buckling where friction mobilization begins saturating during compression. Since small rafts are homogeneous (see \cref{compaction}), the buckling condition simply writes $\phi_0 L_0 = \phi_\text{max}L_\text{Buck}$, 
which yields $\lambda_{u_0} \simeq u_0\times \tfrac{\phi_0}{\phi_\text{max}-\phi_0}$. In the present study, we have $\lambda_{u_0} \simeq 0.73$.
Full force transmission 
is thus achieved for a relatively important range of aspect ratio. Friction is then progressively mobilized for intermediate raft aspect ratio, up to the Janssen limit.

%
For $L_\text{Buck}/W > \lambda_\text{j} \simeq 2.17$, the particles begin accumulating at the compression barrier. A rigidity gradient develops and enhances screening. 
This compensates the partial friction mobilization and drives the observed screening towards exceeding the Janssen limit. As a consequence, we observe the crossing of the Janssen's model and the experimental data at an aspect ratio near $2.5$ in \cref{transmission}.
Above the penetration length $\lambda_\phi \simeq 4.2$, a tail zone of unmoved particles appears 
and sees very little friction mobilization, which explains the apparent saturation of the force transmission (see the experimental data of \citet{Cicuta2009} and \cref{transmission}). 
To reach $99\%$ surface pressure screening at buckling, we find that our raft must have an aspect ratio at buckling of $8.2\lambda_\text{j}$, much larger than the $4.6\lambda_\text{j}$ predicted by the Janssen's model.
%


%

In conclusion, we found that the classical Janssen's analysis is not sufficient to explain the internal mechanics of confined particle rafts. Using deformable ring sensors, we mapped for the first time the internal surface pressure profile during compression up to the buckling onset. We interpreted its continuous development in terms of progressive friction mobilization, and showed that for rafts shorter than $\lambda_{u_0}$, very low friction is observed. For rafts longer than $\lambda_\text{j}$, we found that a gradient of elastic modulus exists within a localized compaction front and enhances stress screening. We expect these competing effects to be especially important for initially loose granular materials, while the Janssen's model recovers validity near the jamming point.
Our results show that  further studies on progressive friction mobilization and heterogeneities are crucial to the description of particles rafts, shedding new light on the mechanics of confined particle-laden interfaces and granular materials.  
 
%

\vskip 6pt

%
%

%
\begin{acknowledgments}
We acknowledge financial support from Fondecyt/Conycit-Chile through the postdoctoral project n$^{\circ}$ 3160152.  F.M. is grateful to Fondecyt/Anillo Act-1410.  We thank G. Lagubeau, F. Vivanco, D. Vella and E. Cl\'ement for enlightening discussions.\\
\end{acknowledgments}

\bibliography{Granular_raft} 
\bibliographystyle{apsrev4-1} 

\end{document}